\documentclass[english,aip,apl,reprint]{revtex4-1}
\usepackage[T1]{fontenc}
\usepackage[utf8]{inputenc}
\usepackage{siunitx,xfrac}
\DeclareSIUnit{\sqrthz}{\hertz\tothe{\sfrac{1}{2}}}
\DeclareSIUnit{\dBm}{dBm}
\DeclareSIUnit{\baud}{baud}

\usepackage{amsmath}
\usepackage{amssymb}
\usepackage{graphicx}
\usepackage[english]{babel}
\usepackage{textcomp}
\usepackage{braket}

\begin{document}

\title{Digital Communication with Rydberg Atoms \& Amplitude-Modulated Microwave Fields}

\author{David H. Meyer}
\email{dihm@terpmail.umd.edu}
\selectlanguage{english}
\affiliation{Department of Physics, University of Maryland, College Park, MD 20742, USA}
\affiliation{U.S. Army Research Laboratory, Sensors \& Electron Devices Directorate, Adelphi, MD 20783, USA}

\author{Kevin C. Cox}
\affiliation{U.S. Army Research Laboratory, Sensors \& Electron Devices Directorate, Adelphi, MD 20783, USA}

\author{Fredrik K. Fatemi}
\affiliation{U.S. Army Research Laboratory, Sensors \& Electron Devices Directorate, Adelphi, MD 20783, USA}

\author{Paul D. Kunz}
\email{paul.d.kunz.civ@mail.mil}
\affiliation{U.S. Army Research Laboratory, Sensors \& Electron Devices Directorate, Adelphi, MD 20783, USA}

\begin{abstract}
Rydberg atoms, with one highly-excited, nearly-ionized electron, have extreme sensitivity to electric fields, including microwave fields ranging from \SI{100}{\mega\hertz} to over \SI{1}{\tera\hertz}. Here we show that room-temperature Rydberg atoms can be used as sensitive, high bandwidth, microwave communication antennas. We demonstrate near photon-shot-noise limited readout of data encoded in amplitude-modulated \SI{17}{\giga\hertz} microwaves, using an electromagnetically-induced-transparency (EIT) probing scheme. We measure a photon-shot-noise limited channel capacity of up to \SI{8.2}{\mega\bit\per\second} and implement an 8-state phase-shift-keying digital communication protocol. The bandwidth of the EIT probing scheme is found to be limited by the available coupling laser power and the natural linewidth of the rubidium D2 transition. We discuss how atomic communications receivers offer several opportunities to surpass the capabilities of classical antennas.
\end{abstract}

\maketitle


Rydberg atoms, created by nearly ionizing one electron of a neutral atom, can be used to create exquisitely sensitive electric field sensors. This is due to the Rydberg atom's dipole moment, $d$, which scales quadratically with the large, principal quantum number $n$, $d \sim e a_b n^2$, where $a_b$ is the Bohr radius and $e$ is the charge of the electron.\cite{gallagher_rydberg_2005}  By probing many atoms at the standard quantum limit,\cite{wineland_squeezed_1994} Rydberg sensors have the potential to reach many orders of magnitude higher sensitivity than traditional electrometers,\cite{fan_atom_2015} and have many other promising capabilities including high dynamic range,\cite{sedlacek_microwave_2012,anderson_optical_2016} SI traceability and self-calibration,\cite{holloway_broadband_2014,simons_simultaneous_2016,holloway_electric_2017} and operation frequency spanning from \si{\mega\hertz}\cite{miller_radio-frequency-modulated_2016} to \si{\tera\hertz}\cite{wade_real-time_2017}. However, little research has explored the application of quantum sensors (like Rydberg atoms) for precise, high-bandwidth classical communication using electro-magnetic fields.\cite{gerginov_prospects_2017}

\begin{figure}[bt]
\begin{centering}
\includegraphics[width=1.0\linewidth]{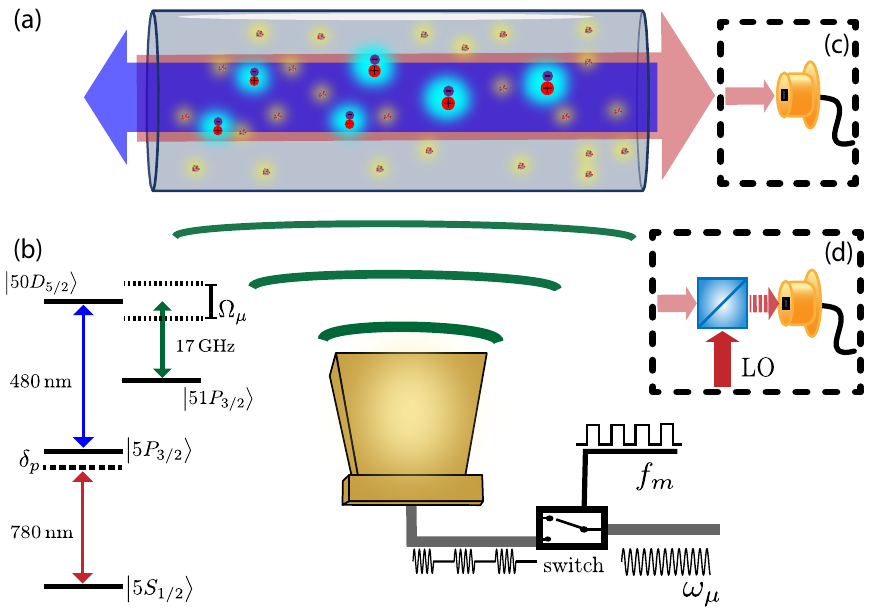}
\par\end{centering}

\caption{\label{fig:Setup}(a) Probe (red) and coupling (blue) light counter-propagate in a vapor cell of rubidium atoms, forming a ladder-EIT system (shown in (b)) that excites ground-state atoms to a Rydberg state. Microwaves (green) from a horn antenna couple the $\ket{50D_{5/2}}$ and $\ket{51P_{3/2}}$ state, and split the EIT peak. A switch modulates the microwaves, and thereby the EIT splitting, and this is detected as amplitude modulation of the probe laser intensity.
(c) Probe intensity modulation can be measured directly with a fast photodetector
(d) or measured using an optical heterodyne, where a local oscillator (LO) beam is mixed with the transmitted probe.}
\end{figure}

\begin{figure*}[bt]
\begin{centering}
\includegraphics[width=1.0\linewidth]{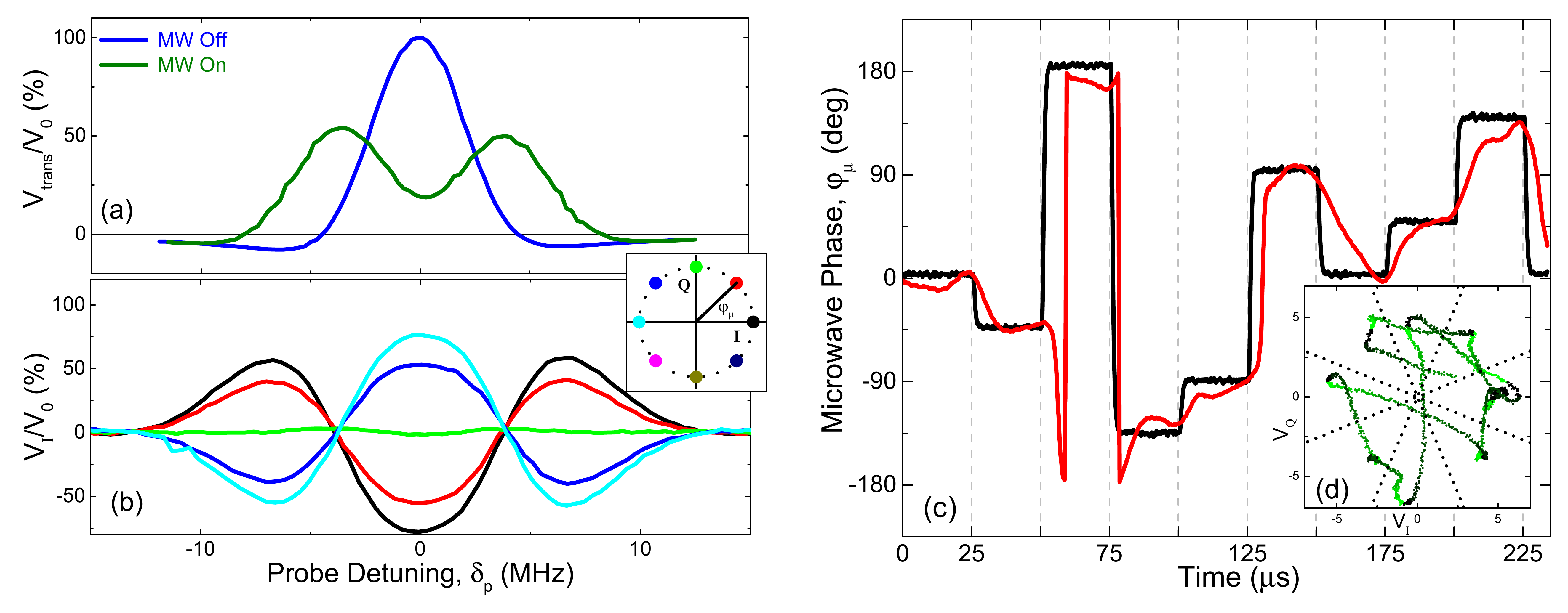}
\par\end{centering}

\caption{\label{fig:CompAndPSK}(a) We observe Rydberg EIT (blue) and Autler-Townes split Rydberg EIT (green) by measuring probe transmission $V_\textrm{trans}$.  
(b) Example demodulated transmission signals $V_\text{I}$ with color corresponding to amplitude modulation phases $\phi_\mu = $ \numlist[]{0;45;90;135;180} degrees, matching the amplitude modulation phase states shown in the inset.  
(c) PSK sent and received phase (black and red, respectively) with $\Omega_{\mu}=2\pi\times\SI{11.4}{\mega\hertz}$ at a \SI{40}{\kilo\hertz} symbol frequency and an amplitude modulation rate of \SI{1.98}{\mega\hertz}. The vertical dashed lines delineate the individual symbol periods. 
(d) Phase constellation of the received phase in (c) (red line). The axes $V_\text{I}$ and $V_\text{Q}$ are in volts measured at the lock-in amplifier. The dashed lines delineate the eight phase states. Marker colors ranging from black to green denote the passage of one \SI{25}{\micro\second} symbol period.}
\end{figure*} 

Rydberg atoms offer several exciting possibilities to exceed what is possible with classical dipole antennas for classical digital communication. First, multiplexing communication using many transitions from \numrange{0.1}{1000} \si{\giga\hertz} may lead to parallel, fast communication in multiple, widely disparate bands. Second, optically-interrogated Rydberg atoms avoid internal thermal noise that can limit classical antennas since the internal states of atoms can be optically pumped to effectively zero-temperature;\cite{bernheim_optical_1965} the readout noise is instead limited by the quantum projection noise of either the probing light (as seen in this work) or, in the ideal case, the atoms. Even when limited by the probing light, Rydberg atoms have already been shown to have record sensitivity down to \SI{0.3}{\milli\volt\per\meter\sqrthz}.\cite{kumar_rydberg-atom_2017} Finally, Rydberg atomic receivers could also, in principle, be used for sub-wavelength imaging\cite{holloway_sub-wavelength_2014,fan_subwavelength_2014,wade_real-time_2017} and vector detection\cite{sedlacek_atom-based_2013}. Recent work has also shown cold Rydberg atoms can mediate direct, coherent electro-optical conversion of MW photons into the optical regime via six-wave mixing.\cite{kiffner_two-way_2016,han_coherent_2018} Given these potential strengths we introduce Rydberg atoms as a new potential platform for digital communication worthy of in-depth study.

In this work, we show that room-temperature Rydberg atoms can be used to implement a microwave-frequency (MW) receiver ``antenna'' 
\footnote{Our driven atomic sensor may not satisfy the definition of an antenna used in classical antenna theory since it breaks several common assumptions. Most importantly our atomic receiver is not passive, and actually performs a non-destructive measurement of the field.} 
for classical, digital communication.  We demonstrate phase-sensitive conversion of amplitude-modulated MW signals into optical signals and perform a demonstration of 8-state phase-shift-keying (PSK), the canonical digital communication protocol.  We also measure the bandwidth limit of our electromagnetically-induced-transparency (EIT) probing scheme and observe a near photon-shot-noise limited channel capacity of up to \SI{8.2}{\mega\bit\per\second} for \SI{395}{\milli\volt\per\meter} microwaves. Finally, we calculate the expected atom-shot-noise limited performance of a Rydberg receiver and find it to be an order of magnitude better than the present measurement.


A schematic of our experimental apparatus is shown in Fig. \ref{fig:Setup}(a), and the core elements are similar to those of other EIT-based Rydberg electrometry work\cite{sedlacek_microwave_2012,holloway_broadband_2014} (see Supplementary Material for additional details).  We use a MW horn to address a \SI{17.0415}{\giga\hertz} transition between the $\ket{50D_{5/2}}$ and $\ket{51P_{3/2}}$ states.  The resonant microwave field establishes an Autler-Townes splitting, proportional to the MW Rabi frequency $\Omega_{\mu}$, which is probed using EIT, shown in Fig. \ref{fig:Setup}(b).  The \SI{480}{\nano\meter} coupling beam counter-propagates with respect to the \SI{780}{\nano\meter} probe to largely cancel Doppler-broadening of the room-temperature atoms.  To observe EIT we either directly measure the transmitted probe power (Fig. \ref{fig:Setup}(c)) or perform heterodyne detection by interfering the probe with a \SI{78.5}{\mega\hertz}-detuned local oscillator (LO).

In Figure \ref{fig:CompAndPSK}(a) we present an example measurement of probe transmission, $V_{\textrm{trans}}$, normalized to the amplitude of the EIT peak $V_0$, observing EIT with the microwaves on (green trace) and off (blue trace).  To send digital information we amplitude modulate the MW field.  The modulation phase $\varphi_\mu$ encodes 8 states, corresponding to all permutations of 3 bits ranging from 000 to 111.  The possible states are shown in the I-Q plane in the inset of \ref{fig:CompAndPSK}(b).  The amplitude modulation is imposed, through EIT, onto the probe laser transmission, and the resulting oscillating probe transmission is then demodulated into an In-Phase voltage $V_\text{I}$ and a Quadrature-Phase voltage $V_\text{Q}$ using a lock-in amplifier. Five distinct examples of the demodulated signal $V_\text{I}$ are plotted versus probe detuning in Fig. \ref{fig:CompAndPSK}(b).  These $V_\text{I}$ signals are proportional to the subtraction of two EIT signals $V_\textrm{trans}$, with microwaves on and off, such as those shown in Fig \ref{fig:CompAndPSK}(a), which leads to features dependent on the MW Rabi frequency $\Omega_{\mu}$ as described in the Supplementary Materials.


We demonstrate a PSK protocol by rapidly changing the phase $\varphi_\mu$ of the amplitude modulation while measuring the lock-in signals at zero detuning. $\varphi_\mu$ is reconstructed from $\varphi_\mu = \arctan(V_\text{Q}/V_\text{I})$.  Figure \ref{fig:CompAndPSK}(c) shows example sent and received amplitude-modulation phases (black and red traces respectively) where each symbol representing three bits of data is transmitted for \SI{25}{\micro\second}. Figure \ref{fig:CompAndPSK}(d) shows the same recovered signal in the corresponding phase space shown in the inset of Fig. \ref{fig:CompAndPSK}(b). Effective signal recovery is done when the demodulation phase is optimized (\emph{i.e.} rotation of the phase space) and the clock is properly recovered (\emph{i.e.} sampling the correct set of data points spaced by the symbol send period). The data transmission rate in this experimental configuration is ultimately limited to $\SI{\sim1}{\mega\bit\per\second}$ by the speed of the lock-in amplifier.\footnote{This is due to the lock-in amplifier's minimum output time constant of $\SI{1}{\micro\second}$ which limits the symbol rate.} In order to show the potential utility of the Rydberg receiver we next characterize the more fundamental limits.


\begin{figure*}[tb]
\begin{centering}
\includegraphics[width=1.0\linewidth]{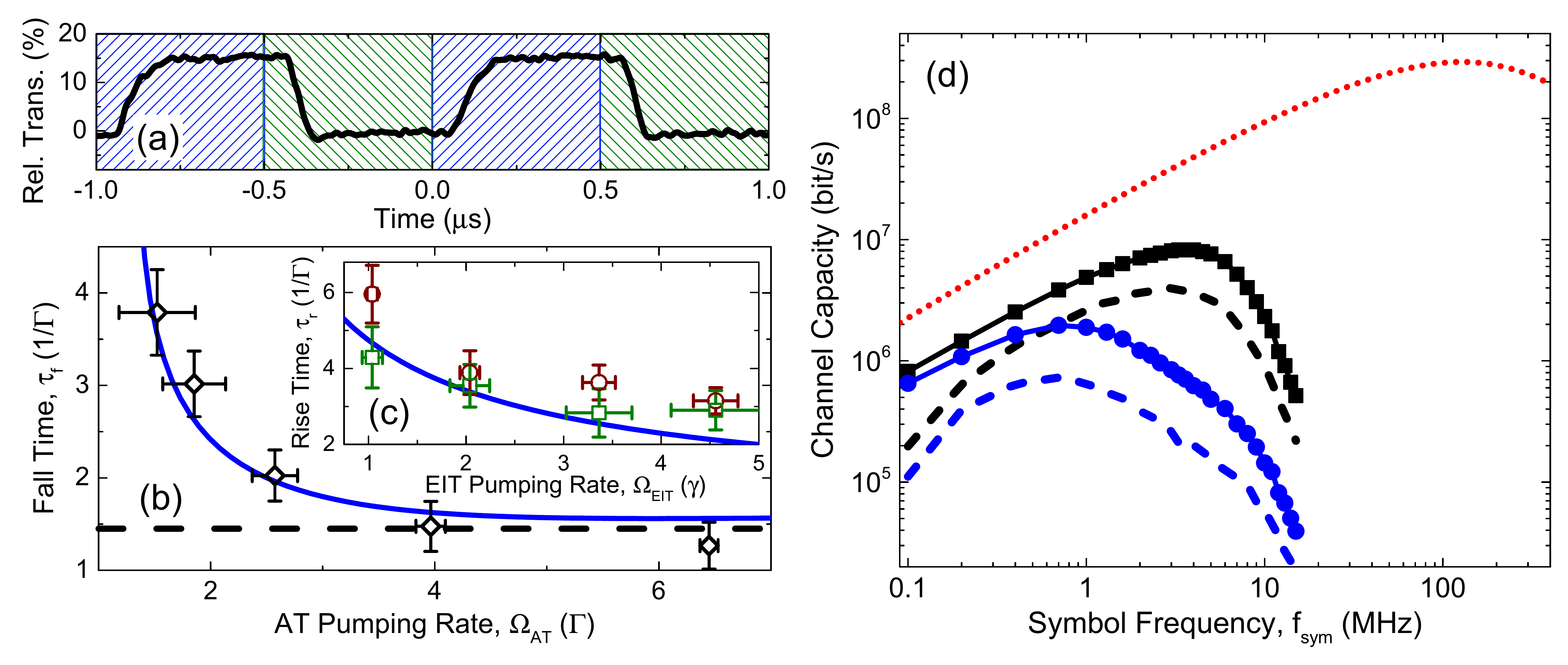}
\par\end{centering}

\caption{\label{fig:Cap}\label{fig:Timetrace}
(a) An example time-domain trace of the transmitted probe signal as MWs are modulated at \SI{1}{\mega\hertz}. The blue and green regions show when the MWs are off and on, respectively.
(b) Measured fall times $\tau_f$ of the time traces versus characteristic pumping rate $\Omega_{AT}$ set by the MW and coupling laser powers.  A numerical model is shown as a solid line, and the dashed black line shows the asymptotic fall time ($1.45/\Gamma$) for pumping into the dark state. 
(c) Measured rise times $\tau_r$ versus the characteristic EIT pumping rate $\Omega_{EIT}$ set by the coupling laser power. The green and maroon points show the measured rise times for the lowest and highest MW powers measured ($\Omega_{\mu}=2\pi\times\SI{5}{\mega\hertz}$ \& $2\pi\times\SI{38}{\mega\hertz}$, respectively). The blue line shows the rise times from the numerical model.
(d) Empirical channel capacity versus symbol frequency with an $\Omega_{\mu}=2\pi\times\SI{8}{\mega\hertz}$ MW Rabi frequency for two coupling powers: $P_c = \SI{48}{\milli\watt}$ (black) and $P_c=\SI{11}{\milli\watt}$ (blue). The dashed lines represent the channel capacity including all noise sources. The lines with symbols represent the channel capacity with only photon-shot-noise considered. The dotted red line shows the theoretically-predicted atom-shot-noise limited capacity for the same MW power.}
\end{figure*}

One of the most important figures of merit for any receiver is the maximum channel capacity $C$ for a single channel. This is given by the Shannon-Hartley Theorem in terms of the signal-to-noise ratio in variance (SNR) and the channel bandwidth (BW), and can be used to determine the achievable communication rate for a channel given a symbol frequency $f_{\text{sym}}$, the measured signal $\mathbb{S}$ in volts, and the voltage noise spectral density $\mathbb{N}$ (assumed here to be random white noise):\cite{hanzo_single-_2000}
\begin{equation}
C=\text{BW}\log_2\left(1+\text{SNR}\right)=f_{\text{sym}} \log_2\left(1+\frac{\mathbb{S}^2}{\mathbb{N}^2f_{\text{sym}}} \right)\textit{.}\label{eq:capacity}
\end{equation}
The channel capacity is optimized when $\text{SNR}=1$ and will, in principle, saturate with increasing $f_{\text{sym}}$ beyond the optimal point. Often the signal degrades with increasing symbol frequency beyond a characteristic frequency determined by the atoms and the measurement method used, leading to a reduction in channel capacity.

We first experimentally measure and theoretically model the bandwidth of our Rydberg receiver. Figure \ref{fig:Timetrace}(a) shows a typical time-domain trace of the signal where the MW field is modulated on and off (green and blue regions respectively) at \SI{1}{\mega\hertz}. Exponential time constants are extracted from the signal time traces and plotted against the relevant pumping rates in Fig. \ref{fig:Timetrace}(b) and (c). This type of measure for the system bandwidth is similar to switching time measures in photon-switching\cite{harris_photon_1998,yan_observation_2001,braje_low-light-level_2003} and cross-phase modulation\cite{schmidt_giant_1996,fleischhauer_electromagnetically_2005} systems based on EIT, which allows us to apply theory developed in those contexts to understand the limitations of our system bandwidth.

The $1/e$ fall times $\tau_f$, corresponding to the MWs being turned on, are shown in Fig. \ref{fig:Timetrace}(b) for five different MW powers and a fixed coupling power $P_{c}=\SI{48}{\milli\watt}$. The MW and coupling Rabi frequencies combine to form the characteristic Autler-Townes pumping rate, $\Omega_{AT}\equiv\sqrt{\Omega_{c}^2+\Omega_{\mu}^2}$, shown on the horizontal axis. The blue lines in Fig. \ref{fig:Timetrace}(b) and (c) show results obtained from numerically integrating the optical Bloch equations for our 4-level system under the same experimental parameters.  We apply a constant multiplicative scale factor to both of these theory lines, corresponding to systematic under-estimation, due to unaccounted-for Doppler-broadening and magnetic sublevels. The value of this parameter, $1.2$, is consistent with similar offset effects described in [\onlinecite{dmochowski_experimental_2016}]. The measured data show good agreement with the numerical model over the entire experimental range. In the weak probe/strong EIT regime ($\Omega_{p}\ll\Gamma$, $\Omega_{c}^2/\gamma\Gamma\gg1$, where $\Gamma$, $\gamma$ are the dephasing rates for the intermediate and ground states) a simple analytical model for laser-cooled atoms, which assumes no dephasing in the ground state, predicts that the fastest possible switching time is $2/\Gamma$.\cite{chen_transient_2004} This model is readily generalized to include ground-state dephasing (dominated by transit effects, as described in the Supplementary Materials) to give a limit of $2/\left(\Gamma+2\gamma\right)=1.45/\Gamma$, with $\gamma=2\pi\times\SI{1.14}{\mega\hertz}$ as the estimated ground-state dephasing rate. The black-dashed line of Fig. \ref{fig:Timetrace}(b) shows this limit. This analytical prediction is also confirmed by our data.

Figure \ref{fig:Timetrace}(c) shows the $1/e$ rise times corresponding to the MWs being turned off, $\tau_r$. This situation represents the well-studied EIT pumping rate, $\Omega_{EIT}\equiv\Omega_{c}^2/2\Gamma$, for a ladder EIT system and translates to the time needed to establish the EIT dark state (i.e. resulting in greater probe transmission).\cite{feizpour_short-pulse_2016} We show fitted rise times for the lowest and highest MW powers (green and maroon points, respectively) versus $\Omega_{\text{EIT}}$ in units of the ground-state dephasing rate $\gamma$. In these units, $\Omega_{EIT}\gg1$ is considered to be in the strong EIT regime. As expected, $\tau_r$ scales inversely with $\Omega_{\text{EIT}}$. We also see reasonable quantitative agreement with the numerical model. In Fig. \ref{fig:Timetrace} parts (b) and (c), for our current parameters, the best achievable rise time $\tau_r$ is slower than the best fall time $\tau_f$, showing we are limited by the coupling power.  However, we note that for a sufficiently strong coupling laser, the fall-time limit would be the bandwidth limit for our EIT probing scheme.

Another consideration that may affect the SNR and bandwidth is dipole-dipole interactions between ground-Rydberg states and between Rydberg-Rydberg states. In this work the low atom density of our vapor cell limits the effects of interactions relative to the Doppler and transit broadening. However, increasing the atom density to improve SNR or going to higher principal quantum numbers to use lower MW frequencies may lead to complications and increased dephasing from collisional broadening or Rydberg blockade due to dipole-dipole interactions.\cite{urvoy_strongly_2015,kara_rydberg_2018}

To measure the photon-shot-noise-limited channel capacity we change the measurement scheme to the heterodyne configuration (see Fig. \ref{fig:Setup}(d)). In heterodyne, gain from the LO amplifies the signal and increases the photon-shot-noise. For sufficiently high LO powers, the photon-shot-noise becomes the dominant noise source, allowing one to disregard other technical noise sources. We record the transmission signal amplitude $\mathbb{S}=V_\textrm{trans}$ and output voltage noise spectrum $\mathbb{N}$ using a spectrum analyzer. Measuring these quantities versus the symbol frequency (assuming one symbol per modulation period) allows us to calculate the maximum attainable channel capacity via Eq. \ref{eq:capacity}. Figure \ref{fig:Cap}(d) shows the empirical channel capacities for our highest (black) and lowest (blue) coupling powers. The dashed lines show the channel capacity including all measured noise sources (\textit{i.e.} detector and laser frequency noise) while the lines with data points show the channel capacity with only photon-shot noise. There are three primary regions of interest. For low symbol frequency the channel capacity is limited by the modulation rate, and shows a linear rise in capacity. The channel capacity then peaks when SNR is reduced to $1$ by the increasing photon-shot noise and decrease in signal, due to the limiting bandwidths $\tau_r$ and $\tau_f$ described above. For higher symbol rates the bandwidth-limited signal reduction dominates and the channel capacity decreases rapidly. 

The maximum empirical channel capacity for the $\Omega_{\mu}=2\pi\times\SI{8}{\mega\hertz}$ MW field (\SI{395}{\milli\volt\per\meter}) shown is \SI{8.2}{\mega\bit\per\second} at a \SI{4}{\mega\hertz} symbol rate. As already described, this capacity is significantly limited by the EIT probing scheme and the associated photon shot-noise. Even so, the sensitivity of the photon-shot-noise limited Rydberg receiver detecting a $\sim\SI{13}{\milli\volt\per\meter}$ MW field allows for a channel capacity of \SI{10}{\kilo\bit\per\second} which is sufficient for some applications such as audio transmission.

However, the photon-shot-noise that is limiting our current measurement is not a fundamental limit. Wave-function collapse limits the SNR when using Rydberg atoms, or any other sensor made of 2-level quantum systems, to the Standard Quantum Limit for measurement of a quantum phase $\phi$. For classical communication the maximum possible accumulated phase for one transmitted symbol is $\phi=\Omega_{\mu}\cdot t_m$, assuming a coherence time longer than the symbol period $t_m=1/f_{m}$. The standard quantum limited phase uncertainty is $\Delta \phi = 1/\sqrt{N}$, where $N$ is the number of qubits (atom number in our case).\cite{wineland_squeezed_1994} This leads to a quantum-limited SNR, $\text{SNR}_{q} = N \Omega_{\mu}^2/f_{m}^2$, and theoretical channel capacity, $C_{q} = f_{m}\log_2\left(1+N\Omega_{\mu}^2/f_{m}^2\right)$. This capacity is optimized when $f_{m} \approx \sqrt{N}\Omega_{\mu}/2$, leading to an optimum quantum-limited capacity of $C^{opt}_q \approx \sqrt{N}\Omega_{\mu}\log_2(5)/2$. The red line of Figure \ref{fig:Cap}(d) shows this predicted channel capacity for the same MW field. The maximum channel capacity for 1000 atoms is predicted to be $\sim\SI{292}{\mega\bit\per\second}$, a more than 30-fold increase over the current maximum attained with photon-shot-noise. Finding a way to achieve the standard quantum limit with room-temperature Rydberg sensors is a significant avenue for further research.  Due to EIT's inherent photon-shot-noise and bandwidth limits that we have characterized here, other probing schemes may be advantageous. Recently realized alternatives that do not rely on optical probing include state-selective ionization of the Rydberg state followed by detection with a multi-channel plate.\cite{facon_sensitive_2016,koepsell_measuring_2017} Electro-optic conversion from MW to optical fields, while different than what we have demonstrated (EIT probing converts only the modulation from MW to optical fields), is another alternative.\cite{han_coherent_2018} However, high efficiency operation requires mode-matching the MW field to the mode volume of the mixing light fields,\cite{kiffner_two-way_2016} which can be difficult for free space communications. 

Notably, the quantum-limited channel capacity does not a function of the carrier frequency.  This is significantly different from the standard data transmission limit for electrically small antennas, often referred to as the Chu limit, which states that the characteristic quality factor (inverse fractional bandwidth) of an electrically-small, efficient antenna is $Q \approx 1/(k^3a^3)$ where $a$ is the size of the antenna and $k$ is the wave-vector of the MW field.\cite{wheeler_fundamental_1947,chu_physical_1948} In contrast, Rydberg atomic sensors can be used with a characteristic size of a few micrometers,\cite{kubler_coherent_2010} while generally maintaining their bandwidth. As a quantitative comparison, consider the 802.11ac Wi-Fi standard for a \SI{5}{\giga\hertz} carrier. This standard has a maximum single channel data rate of \SI{867}{\mega\bit\per\second} at a BW of \SI{160}{\mega\hertz}.\cite{noauthor_ieee_2013} However, if one were to use an efficient, electrically small antenna \SI{500}{\micro\meter} in size (a cell size readily achievable with Rydberg atoms\cite{kubler_coherent_2010}) the BW would be reduced to less than \SI{90}{\kilo\hertz}, significantly reducing the data rate. This comparison is somewhat simplistic; atoms and classical antennas measure electric fields in fundamentally different ways, and an in-depth comparison should be the subject of future work. However, it highlights a potential regime where atomic sensors may significantly outperform a fundamental limit of classical antennas for communication. Rydberg quantum sensors may offer an alternative approach to achieve simultaneously small, high speed, and exquisitely sensitive receivers.  As previously mentioned, they also have many additional attractive properties:  because there are so many available Rydberg states, changing the carrier frequency only amounts to changing the laser frequency, requiring no moving parts; vector detection and sub-wavelength imaging are also possible.  Further study, beyond the scope of this initial work, will be required to discover the precise applications where Rydberg quantum sensors can outperform existing antennas. Nonetheless, this system holds significant promise to become one of a small number of useful technologies operating in the quantum regime.

\section*{Supplementary Material}
See supplementary material for a detailed description of the experimental apparatus, the demodulated signal dependence on MW Rabi frequency, and the derivation of the $\tau_f$ limit for a transit-broadened medium.

\begin{acknowledgments}
This work was partially supported by the Oak Ridge Associated Universities and the Quantum Science \& Engineering Program of the Office of the Secretary of Defense.
\end{acknowledgments}

\bibliographystyle{aipnum4-1}
\bibliography{RydElectrometry}



\setcounter{equation}{0}
\setcounter{figure}{0}
\setcounter{table}{0}
\renewcommand{\theequation}{S\arabic{equation}}
\renewcommand{\thefigure}{S\arabic{figure}}

\onecolumngrid
\begin{center}
\textbf{\large Supplementary Materials for \emph{Digital Communication with Rydberg Atoms \& Amplitude-Modulated Microwave Fields}}
\end{center}
\twocolumngrid
\section*{Experimental Details}

The \SI{780}{\nano\meter} probe light is generated by an external cavity diode laser, linewidth \SI{\sim150}{\kilo\hertz}, that is beat-note locked to a distributed Bragg reflector diode laser stabilized via saturated absorption spectroscopy to a separate reference vapor cell. The probe light power is actively stabilized using an acousto-optic modulator (AOM) and is focused to a $\sfrac{1}{e^{2}}$ radius of \SI{100}{\micro\meter} at the center of a \SI{75}{\milli\meter} long vapor cell. At room-temperature, we measure the optical depth to be $\sim0.4$. The strong \SI{480}{\nano\meter} coupling beam is also focused, to a $\sfrac{1}{e^2}$ radius of \SI{50}{\micro\meter}, so as to have sufficient coupling to the Rydberg state and is generated by a commercial doubling system (Toptica SHG-Pro) that is stabilized to an ultra-low expansion (ULE) reference cavity. This stabilization reduces the linewidth to approximately \SI{2}{\kilo\hertz}. The probe and coupling light are overlapped using a dichroic mirror. These beams counter-propagate in a room temperature, natural abundance rubidium vapor cell with vertical linear polarizations. The microwave (MW) field to be measured is also vertically polarized and propagates perpendicular to the light beams.

The MW field at the resonant \SI{17.0415}{\giga\hertz} frequency is synthesized by a Rhode-Schwarz SMF100A signal generator and applied to the vapor cell using a WR51 waveguide horn antenna. The absolute generator power was calibrated using standard Autler-Townes measurements with MW field Rabi frequencies greater than the electromagnetically-induced-transparency full-width half-max linewidth $(\Gamma_{\text{FWHM}}\approx\SI{4}{\mega\hertz})$. In this regime the MW Rabi frequency $\Omega_{\mu}$ (and by extension the electric field amplitude $|E_{\mu}|$) is related to the Autler-Townes peak splitting $\Delta f$ by
$\Omega_{\mu}=\wp_{\mu}\left|E_{\mu}\right|/\hbar=2\pi D\Delta f$, 
where $\wp_{\mu}$ is the dipole moment of the MW transition.\cite{holloway_broadband_2014} The scaling factor $D=\lambda_{p}/\lambda_{c}$ is due to residual Doppler shifts from the mismatched probe and coupling wavelengths, described in detail below. The generator output is extrapolated from this calibration for $\Delta f<\Gamma_{\text{FWHM}}$ where the AT measurement is no longer resolvable. The MW field modulation is done using an external MW switch (Hittite HMC-C019) on the output of the MW signal generator. The phase of the modulation is controlled via the phase of the TTL-control signal.

The probe intensity modulation due to the amplitude modulated MWs is measured by a fast photodetector (Thorlabs PDA10A) behind a \SI{780}{\nano\meter} laser line filter in two configurations. The first configuration is direct detection (see Fig. 1(c) of the main text) with the signal analyzed using: a lock-in amplifier (Stanford Research SRS865) to demodulate the signal into $I$ and $Q$ quadratures, as done for the data of Fig. 2 (the modulation for 2(a) is on the coupling light, the modulation for 2(b) and (c) on the MWs); or a digital storage oscilloscope (Keysight DSOX1102G) to obtain the time-domain response, as done for the data of Fig. 3(a-c). The second configuration is optical heterodyne detection, in which a strong (\SI{\sim4}{\milli\watt}) local oscillator derived from the probe laser, shifted \SI{78.5}{\mega\hertz}, is interfered with the transmitted probe using a 50/50 fiber splitter. The resulting beat signal is measured using an identical fast photodetector with the output sent to a spectrum analyzer (Agilent N9020A), as done for the results shown in Fig. 3(d).

Overall experimental control and timing is implemented using the open-source labscript suite.\cite{starkey_scripted_2013}

\section*{Derivation of scaling factor $D$}

While always stated in Rydberg electrometry manuscripts,\cite{sedlacek_microwave_2012,holloway_broadband_2014,holloway_electric_2017} the derivation of the scaling factor $D$ in the electrometry context is not readily available in the literature. Here we provide the derivation, which is a specific application of the derivations used in the context of fine and hyperfine splitting measurements using ladder-EIT.\cite{sargsyan_electromagnetically_2010,mack_measurement_2011}

We begin by supposing a ladder-EIT measurement of an AT-split excited state like that shown in Fig. 1(b) of the main text. In order to see the EIT transmission peaks the probe and coupling light must be two-photon resonant for some velocity class of the thermal atoms: $\Delta p(v)+\Delta c(v)=\pm\frac{\Omega_\mu}{2}$, where $\Delta i(v)=\delta_i+\sfrac{v}{c}\left(\omega_{0i}+\delta_i\right)$ is the detuning of the probe or coupling light seen by atoms with velocity $v$ along the light propagation direction, $\omega_{0i}$ is the atomic resonance, and $\delta_i$ is the light detuning from atomic resonance.

If $\Omega_\mu$ is less than the Doppler linewidth of the probing state ($\sim\SI{500}{\mega\hertz}$ for room temperature rubidium) the strongest observed resonance will always occur with atoms having a velocity such that both the coupling and probe and singly resonant. This enforces that $\Delta p(v)=0$ and $\Delta c(v)=\pm\Omega_\mu/2$. Using these relations we can solve for the scaling factor $D$ for when either the probe or coupling light is scanned over the AT splitting.

When the probe is scanned, the coupling light is kept resonant with the atomic transition, $\delta_c=0$. Solving for the probe detuning that achieves resonance with the AT peaks gives $\delta_p^\pm=\left(\frac{\mp\Omega_\mu}{2}\frac{\omega_{0p}}{\omega_{0c}}\right)/\left(1\mp\frac{\Omega_\mu}{2\omega_{0c}}\right)$. Since $\Omega_\mu\ll\omega_{0c}$ for all conceivable MW powers the denominator can be reduced, leading to a measured probe splitting due to the excited state AT splitting of $2\pi\Delta f=\Omega_\mu\sfrac{\omega_{0p}}{\omega_{0c}}$. This gives a scaling factor $D=\lambda_p/\lambda_c$.

When the coupling light is scanned, the probe is kept resonant with the atomic transition, $\delta_p=0$. Again solving for the coupling detunings that acheive resonance with the AT peaks gives $\delta_c^\pm=\pm\Omega_\mu/2$. This leads to a measured AT splitting of $2\pi\Delta f=\Omega_\mu$ and a scaling factor $D=1$.

\section*{Demodulated signal dependence on $\Omega_{\mu}$}

\begin{figure}[tb]
\begin{centering}
\includegraphics[width=1.0\linewidth]{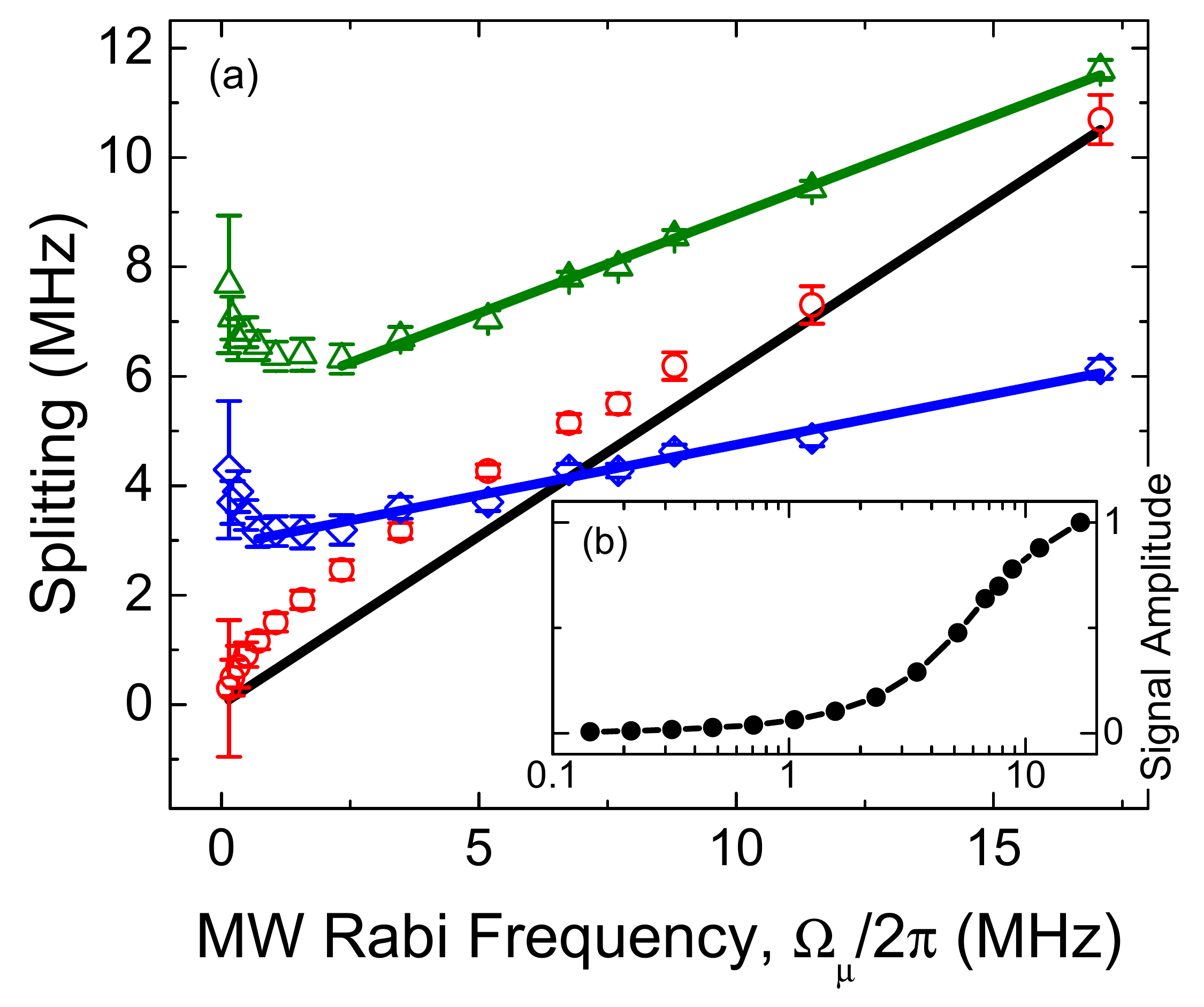}
\par\end{centering}

\caption{\label{fig:FitSummary}(a) Summary of splittings extracted from demodulated signal. The black line is the expected linear AT splitting extrapolated from higher MW field measurements. The red circles are the extracted AT splittings from the Gaussian based model fits. The green triangles show the splitting between the peaks and the blue diamonds show the splitting between zero crossings of the signal. The blue and green lines are linear fits to guide the eye.
(b) Amplitude of demodulated signal with the probe resonant, normalized to the maximum measured signal, versus MW Rabi frequency.}
\end{figure}

In typical Rydberg electrometry measurements the MW field resonantly couples two nearby Rydberg states that are then probed using a ladder EIT scheme. The MW coupling results in Autler-Townes (AT) splitting of the EIT peak that is proportional to the MW Rabi frequency, by $\Omega_{\mu}=2\pi D\Delta f$. This in turn provides a very sensitive, SI-traceable measure of MW electric field amplitudes, as described in the experimental details.\cite{holloway_broadband_2014} However, if the AT splitting is less than the linewidth of the EIT signal the splitting is unresolved and other, less exact, methods must be used.\cite{sedlacek_microwave_2012} Furthermore, even when AT splitting is resolvable, if the splitting is not greater than twice the EIT full-width-half-max linewidth the linear dependence described above is not valid as the atomic system transitions from an AT dominated signal to an EIT dominated signal.\cite{holloway_electric_2017} As a result, techniques for obtaining linear, precise measures of AT splitting near or less than the EIT linewidth are of particular value to the accurate measure of MW electric field amplitudes. The demodulated signal shown in Fig. 2(b) of the main text has two primary features with widths that depend on the MW Rabi frequency that are also clearly resolvable well within the EIT linewidth: the outer peaks splitting and the zero crossings splitting. Furthermore, thinking of the demodulated signal as the simple subtraction of the AT signal by the EIT background hints at the possibility of a way to model the signal to extract the MW Rabi frequency directly. 

Due to residual Doppler-broadening the typical lorentzian peak of EIT becomes approximately gaussian. If we then take the AT peaks to also be gaussian we are able to create a simple model of three gaussian peaks, one with negative amplitude, to fit the demodulated signals and extract the AT splitting. In Fig. \ref{fig:FitSummary}(a) we show the result of these fits compared with the linear AT splitting expected for higher MW powers. In agreement with the results of Ref. [\onlinecite{holloway_electric_2017}] we obtain non-linear deviations from the linear AT splitting when the splitting is less than twice the EIT linewidth ($2\Gamma_{\text{FWHM}}\sim\SI{8}{\mega\hertz}$). This suggests the model is effective at extracting the AT splitting. 

In Figure \ref{fig:FitSummary}(a) we also plot the width of the outer peaks splitting and the zero crossings splitting for the same demodulated signals, the green and blue points respectively. While analytic models of this transitional region are difficult to obtain it is interesting to note that the dependence on the MW Rabi frequency is linear, even when the AT splitting is well within the EIT linewidth. As might be expected, the slope and zero crossings of the linear fits to these data depend on experimental parameters other than the MW field. However, their linear regions continue into the well resolved, AT split regime and therefore could be calibrated precisely using the AT splitting of higher MW fields, effectively extending the regime of linear response to MW fields nearly an order of magnitude lower.

In Figure \ref{fig:FitSummary}(b) we show the scaling of the resonant demodulated signal (shown in Fig. 2(b) of the main text) versus MW Rabi frequency. As can be expected, the scaling of the signal magnitude changes as the signal transitions through the EIT linewidth. This measured scaling allows us to extrapolate the photon-shot-noise limited sensitivity from the measured SNR of the heterodyne measurement in the main text at the relatively high MW field of \SI{395}{\milli\volt\per\meter}. From this we estimate the photon-shot-noise limited sensitivity of the Rydberg receiver to be \SI{0.13}{\milli\volt\per\meter\sqrthz}. Accounting for the differences in experimental parameters, this value agrees with the measured photon-shot-noise limited sensitivity reported by Kumar \emph{et. al.}.\cite{kumar_rydberg-atom_2017} 

\section*{Derivation of $\tau_f$ limit for a Transit-Broadened Medium}

The expression for the fall time $\tau_{f}$, as presented in the footnotes of [\onlinecite{chen_transient_2004}], in the context of laser-cooled atoms, is
\begin{equation}
\tau_f=\frac{2\left(\gamma_{\mu}\Gamma+\Omega_{c}^2+\Omega_{\mu}^2\right)}{\gamma_{\mu}\Omega_{c}^2+\Gamma\Omega_{\mu}^2}\text{.}
\end{equation}
This assumes a typical $N$-level configuration with two excited states with natural linewidths $\Gamma$, $\gamma_{\mu}$ (for the probing and MW transitions, respectively) and the two ground states having infinite lifetime. The four levels are coupled together with optical fields with Rabi frequencies $\Omega_{p}$, $\Omega_{c}$, and $\Omega_{\mu}$ that correspond to the probe, coupling, and MW fields used in our experimental configuration. This result is derived from the optical Bloch equations in the weak probe limit (\emph{i.e.} equations are taken to first order in $\Omega_{p}$ and all population is assumed to be in the lowest ground state.). In the limit of large MW Rabi frequency $\Omega_{\mu}$ the fall time $\tau_{f}$ approaches a minimum value of $2/\Gamma$. Assuming sufficiently large $\Omega_{c}$ such that the EIT condition is met, this time sets the basic bandwidth limit for EIT probing of the MW modulation. 

\begin{figure}[tb]
\begin{centering}
\includegraphics[width=1.0\linewidth]{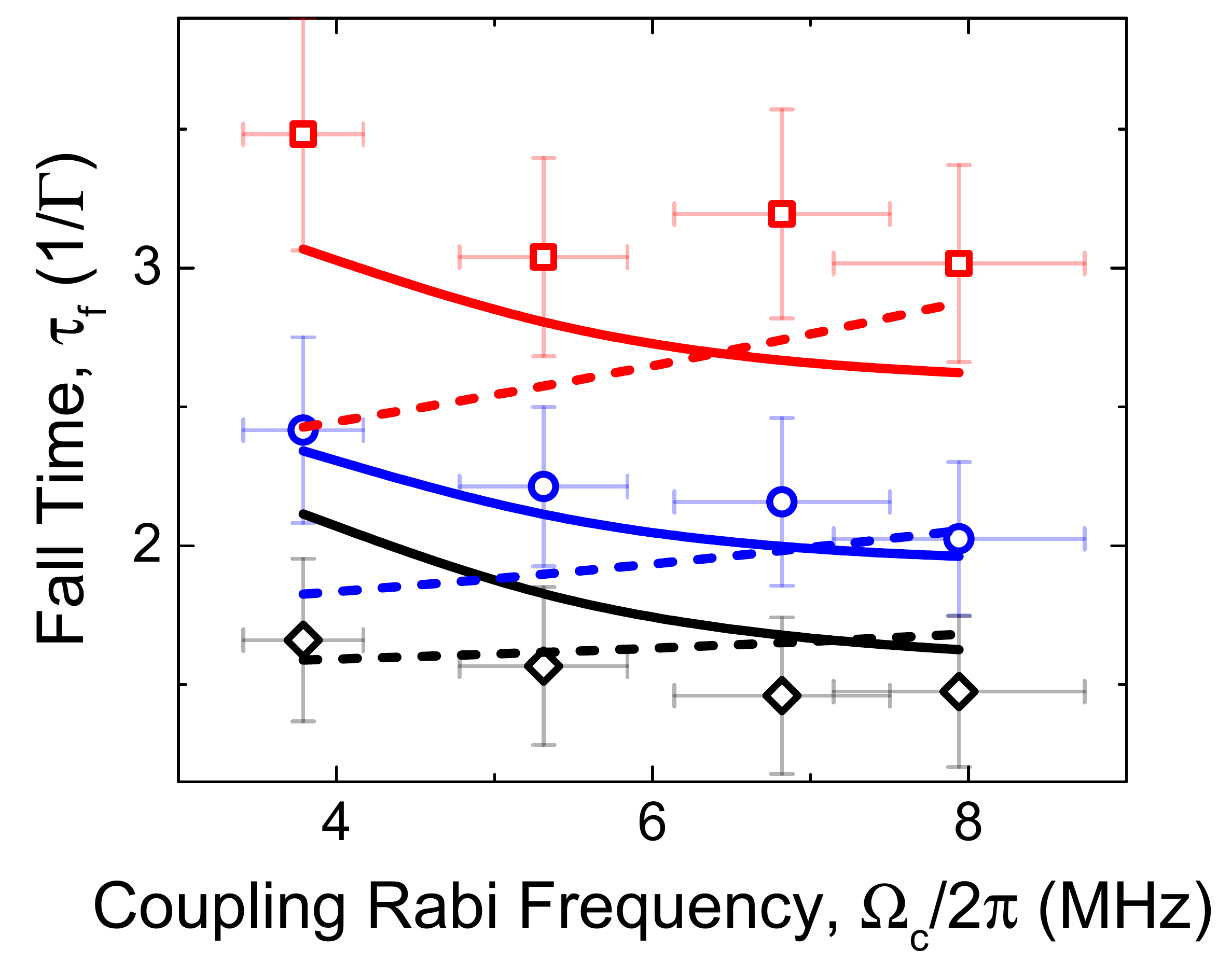}
\par\end{centering}

\caption{\label{fig:FallTimeModels}Comparison of experimental data to numerical and analytical model predictions for the fall time $\tau_{f}$. The red squares, blue circles, and black diamonds show the measured $\tau_{f}$ versus coupling Rabi frequency $\Omega_{c}$ for $\Omega_{\mu}/2\pi=8.0$, $13.4$, and $\SI{22.7}{\mega\hertz}$. The solid lines show the numerical model fits (scaled by $1.2$ as described in the main text) and the dashed lines the analytical model predictions for the same experimental parameters. The data points at $\Omega_{c}/2\pi=\SI{8}{\mega\hertz}$ correspond to the middle data points of Fig. 3(b) in the main text.}
\end{figure}

For warm atoms, as used in this work, ground-state dephasing due to transit effects must also be considered. Transit dephasing is the result of thermal atoms traversing the probe beam profile. While within the profile, an atom interacts with the light as expected and becomes polarized according to the Hamiltonian of the system. Once the atom leaves the profile, it is replaced by a fresh, unpolarized atom, which dephases any coherent effects established. The transit dephasing is related the the probe beam size and the temperature of the atoms. We estimate the total ground-state dephasing, including non-transit sources, to be $\gamma\approx2\pi\times\SI{1.14}{\mega\hertz}$ for our experimental setup.\cite{sagle_measurement_1996}

Under the weak probe approximation, using the method of [\onlinecite{chen_transient_2004}] explained above, and including transit dephasing with all excited-state natural lifetimes for our experimental system, the optical Bloch equations for the coherences induced by the probe, coupling, and MW fields are
\begin{align}
\frac{d}{dt}\rho_{21}&=-\frac{1}{2}\left(\Gamma+2\gamma\right)\rho_{21}+\frac{i}{2}\left(\Omega_{p}+\Omega_{c}\rho_{31}\right)\\
\frac{d}{dt}\rho_{31}&=-\frac{1}{2}\left(\Gamma_{D}+2\gamma\right)\rho_{31}+\frac{i}{2}\left(\Omega_{c}\rho_{21}+\Omega_{\mu}\rho_{41}\right)\\
\frac{d}{dt}\rho_{41}&=-\frac{1}{2}\left(\Gamma_{P}+2\gamma\right)\rho_{41}+\frac{i}{2}\Omega_{\mu}\rho_{31}
\end{align}
where $\rho_{i1}$ is the density matrix element representing the coherence between the $i=(2,3,4)=\left(\ket{5P_{3/2}},\ket{50D_{5/2}},\ket{51P_{3/2}}\right)$ states and the ground state $\ket{5S_{1/2}}$. $\Gamma_{D,P}$ are the natural lifetimes of the $D$ and $P$ Rydberg states. The probe absorption is proportional to $Im(\rho_{21})$ so solving the above equations to obtain an approximate, first-order differential equation for $\rho_{21}$ allows us to model the expected exponential decay of the absorption as the MW field is turned on.

Taking $\Gamma_{D,P}\ll\gamma$, the resulting differential equation becomes
\begin{equation}
\frac{d}{dt}\rho_{21}=-\frac{1}{\tau_{f}}\left(\rho_{21}-\rho_{21}^{(SS)}\right)
\end{equation}
where the steady state value (proportional to the observed transmission) with MWs on is
\begin{equation}
\rho_{21}^{(SS)}=\frac{i\Omega_{p}\left(4\gamma^2+\Omega_{\mu}^2\right)}
{8\gamma^3+4\gamma^2\Gamma+2\gamma\Omega_{AT}^2+\Gamma\Omega_{\mu}^2}
\end{equation}
and the $1/e$ fall time is
\begin{equation}
\tau_{f}=\frac{2\left(12\gamma^2+4\gamma\Gamma+\Omega_{AT}^2\right)}
{8\gamma^3+4\gamma^2\Gamma+2\gamma\Omega_{AT}^2+\Gamma\Omega_{\mu}^2}\text{.}
\end{equation}
When $\Omega_{\mu}\gg\Omega_{c},\Gamma,\gamma$ this reduces to a minimum fall time of $2/\left(2\gamma+\Gamma\right)$, which is approximately $1.45\Gamma$ using the experimental parameters of Fig. 3(b) in the main text. We see that transit dephasing allows us to exceed the cold-atom EIT result ($2/\Gamma$) by providing a second loss mechanism for the atomic coherence established in EIT. In much the same way a parallel resistance can improve the bandwidth of a classical antenna at the expense of signal efficiency, transit dephasing can improve the bandwidth of the EIT probing scheme at a cost of reduced signal.

In Figure \ref{fig:FallTimeModels} we compare this model to experimental fall time data and the numerical model described in the main text versus coupling Rabi frequency $\Omega_{c}$ for a few fixed MW powers. For the higher coupling powers we see good agreement with the measured data. However, the analytical model predicts that the fall time should decrease with decreased $\Omega_{c}$. This expected behavior has been seen in similar EIT-based cross-phase modulation systems where the EIT linewidth, which is inversely proportional to $\Omega_{c}$, sets the bandwidth of the modulation.\cite{feizpour_short-pulse_2016,dmochowski_experimental_2016} However, both the data and the numerical model show the opposite trend. This is due to a breakdown of the analytical model as the weak probe/strong EIT regime assumption becomes less valid. As mentioned in the main text, the EIT regime is when $\Omega_{c}^2/\Gamma\gamma\gg1$ which is only approximately true for the data presented. In this weak EIT regime greater $\Omega_{c}$ leads to a stronger EIT signal that effectively has a larger linewidth and therefore bandwidth. 

In short, increased coupling power to fully obtain the EIT regime is necessary for greater signal bandwidth. However, increasing the coupling power well beyond the EIT regime will eventually lead to reduced signal bandwidth as the EIT linewidth is narrowed. 



\end{document}